\documentclass{article}
\pdfoutput=1

\usepackage{arxiv}

\usepackage[utf8x]{inputenc} 
\usepackage[T1]{fontenc}    
\usepackage{hyperref}       
\usepackage{url}            
\usepackage{booktabs}       
\usepackage{amsfonts}       
\usepackage{nicefrac}       
\usepackage{microtype}      
\usepackage{lipsum}
\usepackage{xcolor}
\usepackage{graphicx}
\usepackage{setspace} 
\usepackage{amsmath}
\usepackage{array}
\usepackage{mdwmath}
\usepackage{mdwtab}
\usepackage{eqparbox}
\usepackage{url}
\usepackage{multirow}
\usepackage{textcomp}
\usepackage{diagbox}
\usepackage{amsmath,amsfonts,amssymb}

\usepackage{caption}
\usepackage{subcaption}

\usepackage{comment}

\title{Non-parametric Bayesian mixture model to study adverse events of COVID-19 vaccines}
\date{March 24, 2023}

\author{
  Ali Turfah\\
  Department of Biostatistics\\
  University of Michigan\\
  Ann Arbor, MI 48109 \\
  \texttt{aturfah@umich.edu} \\
  \And 
  Xiaoquan Wen\thanks{Both Xiaoquan Wen and Lili Zhao are corresponding authors.} \\
  Department of Biostatistics\\
  University of Michigan\\
  Ann Arbor, MI 48109 \\
  \texttt{xwen@umich.edu} \\
  \And 
  Lili Zhao$^*$ \\
  Research Institute\\
  Corewell Health's Beaumont Hospital\\
  Royal Oak, MI 48073 \\
  \texttt{lili.zhao@corewellhealth.org} \\
}

\begin{document}
\maketitle

\begin{abstract}
The vaccine adverse event reporting system (VAERS) is a vital resource for post-licensure vaccine safety monitoring and has played a key role in assessing the safety of COVID-19 vaccines. However it is difficult to properly identify rare adverse events (AEs) associated with vaccines due to small or zero counts. We propose a Bayesian model with a Dirichlet Process Mixture prior to improve accuracy of the AE estimates with small counts by allowing data-guided information sharing between AE estimates. We also propose a negative control procedure embedded in our Bayesian model to mitigate the reporting bias due to the heightened awareness of COVID-19 vaccines, and use it to identify associated AEs as well as associated AE groups defined by the organ system in the Medical Dictionary for Regulatory Activities (MedDRA) ontology. The proposed model is evaluated using simulation studies, in which it outperforms baseline models without information sharing and is applied to study the safety of COVID-19 vaccines using VAERS data.
\end{abstract}



\section{Background}

As of January 1, 2023, COVID-19 vaccines have been administered to over 600 million people in the United States \cite{mathieu2021global}. However, concern about the safety of the vaccines is among the primary factors contributing to vaccine hesitancy \cite{troiano2021vaccine}. There are three COVID-19 vaccines approved by the US Food and Drug Administration (FDA) under emergency use authorization (EUA) in the United States: two mRNA-based vaccines---the BNT162b2 (Pfizer–BioNTech) and the mRNA-1273 (Moderna) vaccines---and the Ad26.COV2.S (Johnson \& Johnson–Janssen) vaccine. 

Due to limited sample size, restrictive inclusion criteria, and the limited duration of follow-up in phase-III trials, a chief concern is that rare or serious adverse events associated with the COVID-19 vaccine have not yet been identified. In addition, despite being theoretically safe, the two mRNA-based vaccines (BNT162b2 and mRNA-1273) are coming to the market for the first time; their safety can only be verified by long-term longitudinal surveillance.

The Centers for Disease Control and Prevention (CDC) and the U.S. Food and Drug Administration (FDA) conduct post-licensure vaccine safety monitoring using the Vaccine Adverse Event Reporting System (VAERS) \cite{shimabukuro2015safety}. VAERS accepts spontaneous reports of suspected vaccine adverse events after administration of any vaccine licensed in the United States from 1990 to the present. As a national public health surveillance resource, VAERS is key for ensuring the safety of vaccines. We define an adverse event (AE) signal as a potential vaccine safety problem, indicating that an AE might be caused by vaccination and warrants further investigation or action. Several AE signals for COVID-19 vaccines have been identified using VAERS data, including anaphylaxis and myocarditis after receipt of mRNA-based COVID-19 vaccines \cite{shimabukuro2021reports, oster2022myocarditis}, and Guillain-Barré syndrome and Cerebral Venous Sinus Thrombosis (CVST) after receipt of the Ad26.COV2.S vaccine \cite{shimabukuro2021reports,woo2021association}. 

As a passive surveillance system, VAERS lacks information on the total number of subjects vaccinated, the total number of vaccine recipients who experience an AE, or the incidence of AEs in unvaccinated subjects. However, the proportion of reports mentioning a specific AE for the COVID-19 vaccines can be compared to the proportion of reports mentioning the same AE for a reference vaccine---or all other vaccines (i.e., non-COVID-19 vaccines). This type of approach is called disproportionality analysis. VAERS data is commonly represented by a large contingency table for the counts of all vaccine and AE pairs over a specified time period. Each row of this table represents a vaccine and each column represents an AE. Each cell in the table contains the number of VAERS reports that mention both the vaccine and the AE during the time period. This table can be very large, with hundreds of rows for vaccines and thousands of columns for AEs. To study the relationship between COVID-19 vaccines and each AE, the large table can be collapsed into a $2\times 2$ table with rows corresponding to reciept of a COVID-19 vaccine (Yes/No) and two columns for the occurrence of the AE (Yes/No). The strength of the vaccine-AE association can be measured using a ratio, such as a proportional reporting ratio (PRR) or an reporting odds ratio (ROR) \cite{zhao2020improvement}. A larger ratio ($>1$) corresponds to a stronger association, which might indicate a vaccine safety problem.

Many disproportionality methods have been used to conduct safety studies based on the VAERS database \cite{DuMouchel:1999, Evans:2001, huang2011likelihood, huang2014review, van:2002, Bate:1998, Orre:2000, Noren:2006, DuMouchel:2001, Szarfman:2002, Kulldorff:2011, Davis:2005, Li:2009, Li:2014, Kulldorff:2013}. Generally these methods investigate each AE separately, assuming that the AE counts follow a Poisson or Binomial distribution. As we know, AEs from vaccines are often rare with sparse count data. In these separate analyses, the ratio parameter for the rare AEs either cannot be estimated or will take an extreme value with a large degree of uncertainty. Therefore, efficient approaches are needed to improve estimation for rare AEs. The Gamma-Poisson Shrinker (GPS) model \cite{DuMouchel:1999} enhances the simple use of the separate AE analyses by allowing information borrowing between similar rate estimates, thereby reducing the sampling variation in the data and improving accuracy for the rare AE estimation. GPS assumes that report counts for the AEs follows a Poisson distribution with a rate representing the strength of the association (i.e., rate $>1$ for a signaled AE). This approach uses a mixture of two distributions to model the rates of  ``signaled"  and ``non-signaled" AEs, assuming that majority of the rates are clustered at or below one (``non-signaled" AEs) while the remaining rates are clustered at a rate above one (``signaled" AEs). Although theoretically sound, it is practically restrictive to approximate the distribution of the rates using only two distributions. The Multi-item Gamma-Poisson Shrinker (MGPS) \cite{DuMouchel:2001} extends the GPS model to adjust for patient risk factors. However, MGPS can only handle categorical risk factors and has increased difficulty estimating the rates for rare AEs because it creates separate contingency tables for each risk factor level combination, exacerbating the sparse counts already present in VAERS.


 In this paper, we estimate the ratio parameter in a regression framework, allowing for both categorical and continuous risk factors. We approximate the distribution of the ratio parameters by a mixture of distributions where the number of mixing components is automatically determined from the data. Specifically, we assume that AE count data follow a binomial distribution, modeling the AE count as a function of the vaccine group (e.g., COVID-19 vaccines Yes/No) and risk factors (such as age and gender) using Bayesian logistic regression. We assign a Dirichlet process mixture (DPM) \cite{ohlssen2007flexible} prior to the log odds ratio parameter; we refer to it as the log reporting odds ratio (logROR) parameter in the VAERS data analysis. The DPM prior uses a mixture of normal distributions to estimate logROR without determining the number of mixing components beforehand; it efficiently accommodates signalled and non-signalled AEs and allows for a more flexible distribution of ROR parameters with the possibility of skewness and multimodality. The logROR parameters with similar magnitudes (within a cluster) will be shrunk towards each other, thereby borrowing information from each other to improve the accuracy of logROR estimates, increasing the power to detect true signals while reducing false positives.


Like all spontaneous reporting systems, VAERS data is vulnerable to reporting bias \cite{zhao2020improvement}. Due to the heightened awareness around the COVID-19 vaccines, AEs observed after administration of COVID-19 vaccines are more likely to be reported than those from other vaccines, which will bias association estimates and result in false positive associations. In order to mitigate this bias, negative controls (NCs)---AEs that are known not to be causally related to any vaccine  \cite{shixu,schuemie2014interpreting}---are often used, requiring the ROR estimates for AE signals to be significantly larger than the ROR estimates of the NCs. Moreover, based on  MedDRA (Medical Dictionary for Regulatory Activities) \cite{Brown:1999}, the largest AE ontology resource describing disease relationships, we conduct enrichment analysis to identify enriched AE groups where signaled AEs occur more frequently compared to other groups. The proposed methods have been implemented in JAGS
and the source code is included in Supplementary Document A.

The rest of the article is organized as follows. In Section 2 we develop the DPM model. In Section 3 we present the results of our simulation studies. In Section 4 we apply the proposed methods to study the safety of COVID-19 vaccines using the VAERS database. Concluding remarks are given in Section 5.


\section{Methods}

\subsection{Bayesian Logistic Regression} 
Suppose there are a total of $N$ reports in VAERS and $J$ AEs mentioned in these reports. Let $V_i \in \{0,1\}$ indicate the type of the vaccine (1 for COVID-19 vaccines and 0 for non-COVID-19 vaccines) for report $i$ $(i = 1,\ldots, N)$. Let $A_{ij}$ be the indicator of mentioning AE $j$ $(j = 1,\ldots, J)$ in report $i$, i.e., $A_{ij} = 1$ if AE $j$ is mentioned in report $i$ and 0 otherwise. Let $\textbf{X}_{i}$ be a vector of covariates for report $i$ (including an intercept term). We use a logistic regression to estimate the strength of association between COVID-19 vaccines and  AE $j$,
$$
logit(Pr(A_{ij}=1|\mathbf{X}_{i}, V_{i})) = \boldsymbol \alpha_j^{\intercal} \mathbf{X}_{i} + \beta_j V_{i}, ~~~ i=1, \cdots, N,$$

where $\beta_j$ is the log reporting odds ratio (logROR) of AE $j$ when comparing a COVID-19 vaccine to a non-COVID-19 vaccine, which is the parameter of interest.




To allow information borrowing for estimating the logROR, we assign  $\beta_j$ a Dirichlet Process Mixture (DPM) of normals prior. That is,
\begin{align*}
    \beta_1, \cdots, \beta_J & \sim G \\
    G & \sim DP(\gamma, G_0),
\end{align*}
where $G_0$ corresponds to a best guess for $G$ as a priori and $\gamma$ expresses confidence in this guess. The stick-breaking representation \cite{sethuraman1994constructive} implies that $G \sim DP(\gamma, G_0)$ is equivalent to
\begin{align*}
    G & = \sum_{k=1}^\infty \pi_k \textbf{N}(\mu_k, \tau_k) \\
    & ~~~~~~~~~~(\mu_k, \tau_k)  \sim G_0 \\
    & ~~~~~~~~~~ \sum_{h=1}^\infty \pi_h  = 1
\end{align*}
The stick-breaking approach parametrizes the component probabilities $\pi_k$ as the proportion of remaining probability after considering $\sum_{h=1}^{k-1} \pi_h$. The proportion of the remaining probabilities for each component is assigned a $Beta(a_0, b_0)$ prior. Recent research has focused on using the constructive definition of the Dirichlet process to produce practical MCMC algorithms \cite{ishwaran2000markov}. The principle is to approximate the full process by truncating the DPM at a maximum number of components $K$, so that
$$
G = \sum_{k=1}^K \pi_k \textbf{N}(\mu_k, \tau_k)
$$
While a large $K$ provides an accurate approximation to the full DPM, it computationally expensive to fit. Strategies have been proposed to specify $K$ \cite{ohlssen2007flexible, ishwaran2000markov}. We found that $K=5$ worked very well for the VAERS data in our applications.

These $K$ components can be interpreted as clusters from which the $\beta_j$ of the AEs are drawn \cite{ohlssen2007flexible}. These clusters enables us to shrink very extreme estimates of $\beta_j$ to more reasonable values as well as obtain more stable estimates for AEs observed in a very small number of cases, neither of which would be feasible if the AEs were to be modeled separately. 

We then specify the hyperpriors for $\mu_k$ and $\tau_k$ in the $k^{th}$ component of the DPM based on \cite{ohlssen2007flexible, halfCauchy},
\begin{align*}
    \mu_k & \sim N(\mu_{base}, \tau_{base}) \\
    \mu_{base} & \sim N(0, \tau_{0}) \\
    \tau_{base} & = \frac{1}{u^2} \\
    u & \sim Uniform(0, f_0) \\
    \tau_k & \sim Gamma(r_\tau, \lambda) \\
    \lambda & \sim Gamma(r_\lambda, \lambda_0),
\end{align*}
where $\tau_0, f_0, r_\tau, r_\lambda, \lambda_0$ are hyperparameters set a priori. For other coefficients $\boldsymbol\alpha_{j}$ in the logistic regression model, we use normal priors with mean 0 and precision $\tau_\alpha$. 


\subsection{Bayesian Negative Control Approach to Identify AE Signals}
In order to mitigate the reporting bias from analyses using spontaneous reporting systems such as VAERS, we identify $36$ AE terms in VAERS as negative controls (NCs). The list of NCs used in our analyses can be found int Supplementary Document A. All of the NCs are known not to be causally related to any vaccine and are reported at least 300 times in our dataset. The chosen NCs occur between 300-16,300 times in our dataset. If there is no reporting bias, the RORs of the NCs should be centered around one. With the aforementioned over-reporting issue, RORs for these NCs will be centered around a value larger than one. Therefore, we can not identify signal AEs based solely on whether their ROR is larger than one or not. In \cite{schuemie2014interpreting} the authors use ROR estimates from the NCs to model the null distribution and use this empirically estimated distribution to yield properly calibrated p-values.

Our Bayesian model can easily incorporate these NCs to make proper posterior inference. At each MCMC iteration, we obtain the logROR estimate for AE $j$ ($\beta_j$) and compare it to the logROR estimates for all NC AEs in that iteration ($\beta_{NC_{1}}, \cdots, \beta_{NC_{36}}$). If $\beta_j$ is larger than majority of the NC logROR estimates ($>35$ NCs is used in this paper) then it is assigned a signal indicator of 1 for the iteration. Otherwise it is assigned a signal indicator of 0 for the iteration. Across all iterations, We compute the posterior probability that AE $j$ is a signal based on the proportion of iterations in which the AE was assigned a signal indicator of 1. Finally, we define an AE as a signal if this NC-adjusted posterior probability is larger than a threshold---a 90\% cutoff is used in this paper.

\subsection{Bayesian Enrichment Test to Identify Enriched AE Groups}

The MedDRA ontology \cite{Brown:1999} groups AEs into System Organ Classes (SOC) based on aetiology, manifestation site, or purpose. For instance, ``Dyspnea", ``Atrial Thrombosis", and ``Myocardial Oedema" are all members of the ``Cardiac Disorder" SOC group. In addition to identifying individual AEs associated with the COVID-19 vaccine, we are interested in identifying SOC groups which are associated with the vaccine; for example, it is important to know that COVID-19 vaccines are associated with cardiac disorders.



Let $\mathcal{S}$ denote the set of all SOC groups. To conduct the enrichment analysis for a particular SOC group, $s$, a common approach is to create a $2\times 2$ table similar to the one presented in Table \ref{tab:soc_enrichment}. The elements of this table are the counts of the $J$ AEs grouped based on their membership in group $s$ as well as their signal AE status (determined based on a significance test). If there are significantly more signaled AEs in $s$, based on a Chi-square or Fisher's exact test, then group $s$ is enriched.

\begin{table}[ht]
    \centering
    \footnotesize
    \begin{tabular}{|c|cc|c|}
    \hline
        \diagbox{SOC Group}{\# AEs} & Signal & Non-Signal & Total \\
    \hline
         $s$ (e.g. Cardiac Disorders) & $a$ & $J_s - a$ & $J_s$ \\
         $\mathcal{S} - s$ (All others) & $c$ & $J_{s^*} - c$ & $J_{s^*}$ \\
    \hline
         Total & $a+c$ &  $J - a - c$ & $J$\\
    \hline
    \end{tabular}
    \caption{Contingency table for SOC group enrichment analysis}
    \label{tab:soc_enrichment}
\end{table}

In this paper, we propose an enrichment analysis in a Bayesian framework. We construct this $2\times 2$ table in every MCMC iteration and perform the enrichment analysis based on the resulting posterior distribution.  Specifically, at each MCMC iteration we obtain the signal indicators for AEs within group $s$ ($a$ signals from $J_s$ AEs) as well as for AEs from all other groups ($c$ signals from $J_{s^*}$ AEs). We then compute an odds ratio (OR) based on Table \ref{tab:soc_enrichment}. This OR represents an enrichment score (we refer to it as the enrichment OR, or EOR); a larger EOR ($>1$) indicates stronger evidence that $s$ is enriched. Across all MCMC iterations, we obtain the posterior distribution of the EOR for $s$. Group $s$ is considered enriched if the lower bound of the 95\% posterior credible interval for the EOR is greater than 1 and the posterior mean of the EOR is greater than a pre-specified value to ensure clinical relevance. 

Compared to the conventional approach described above, our method takes into account the uncertainty in the measurement of signal significance of the AEs; that is, signal indicators may vary across MCMC iterations while the conventional approach has the signal indicators fixed in the enrichment analysis.  Moreover, it is very simple to implement in the Bayesian framework since the test is constructed using by-products of posterior draws.

\section{Simulation Study}

\begin{table}[ht]
\centering
\begin{tabular}{lrrrcrrrcrrr@{\extracolsep{\fill}}}
  \hline
  & \multicolumn{3}{@{}c@{}}{$\sigma = 0.5$} && \multicolumn{3}{@{}c@{}}{$\sigma = 0.8$} && \multicolumn{3}{@{}c@{}}{$\sigma = 1$} \\
\cline{2-4} \cline{6-8} \cline{10-12}
 & DIC & MSE & Coverage && DIC & MSE & Coverage && DIC & MSE & Coverage \\ 
  \hline
  DPM & 3,653 & 0.87 & 95.3\% &&
    3,624 & 1.16 & 93.4\% &&
    3,673 & 0.97 & 96.6\% \\
  IL-Informative & 3,669 & 0.93 & 95.7\% &&
    3,649 & 1.21 & 94.7\% &&
    3,675 & 1.05 & 95.0\% \\ 
  IL-Vague & 3,691 & 6.43 & 94.7\% &&
    3,652 & 6.64 & 94.7\% &&
    3,694 & 6.29 & 96.6\% \\ 
   \hline
\end{tabular}
\caption{DPM and IL model performance in simulation studies.}
\label{tab:simul_results}
\end{table}

\begin{figure}[ht]
\centering
     \hfill
     \begin{subfigure}[c]{0.25\linewidth}
        \centering
        \includegraphics[width=\textwidth]{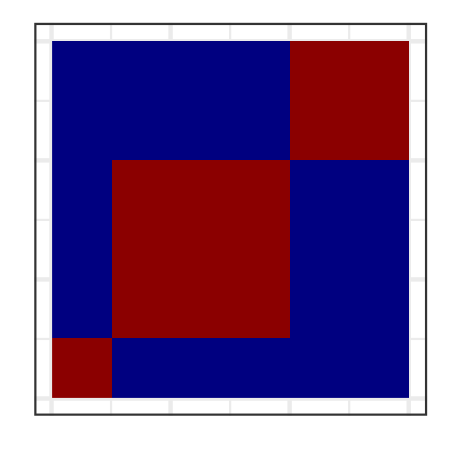}
        \caption{Ground Truth $\sigma = 0.5$ }
        \label{fig:sig05_gt_htmp}
     \end{subfigure}
     \hfill
     \begin{subfigure}[c]{0.25\linewidth}
        \centering
        \includegraphics[width=\textwidth]{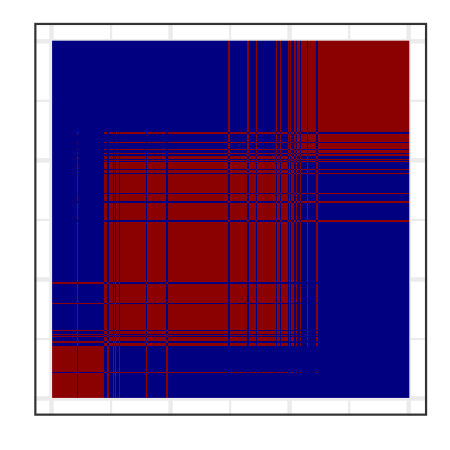}
        \caption{Ground Truth $\sigma = 0.8$ }
        \label{fig:sig08_gt_htmp}
     \end{subfigure}
     \hfill
     \begin{subfigure}[c]{0.25\linewidth}
        \includegraphics[width=\textwidth]{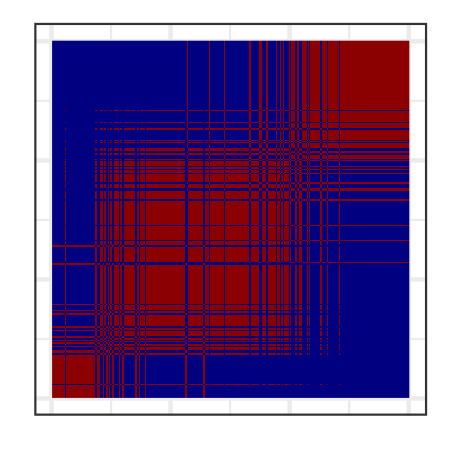}
        \caption{\small{Ground Truth $\sigma = 1$ }}
        \label{fig:sig1_gt_htmp}
     \end{subfigure}
     \hfill
      \begin{subfigure}[c]{0.13\linewidth}
         \centering
        \includegraphics[width=\textwidth]{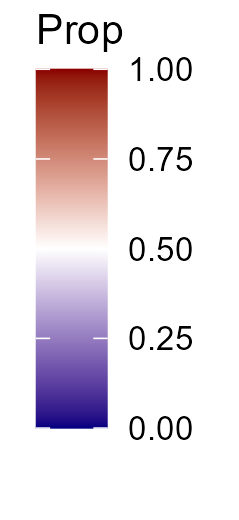}
     \end{subfigure}
     \hfill
     \vskip\baselineskip
    \hfill
     \begin{subfigure}[c]{0.25\linewidth}
        \centering
        \includegraphics[width=\textwidth]{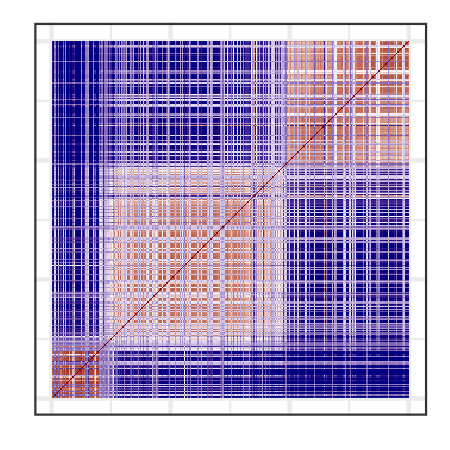}    
        \caption{\small{DPM $\sigma = 0.5$}}
     \end{subfigure}
    \hfill
     \begin{subfigure}[c]{0.25\linewidth}
        \centering
        \includegraphics[width=\textwidth]{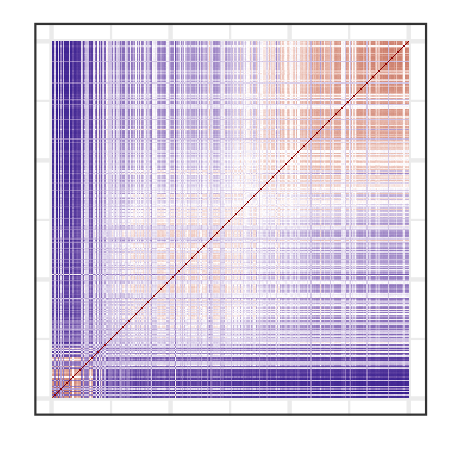}    
        \caption{\small{DPM $\sigma = 0.8$}}
     \end{subfigure}
    \hfill
     \begin{subfigure}[c]{0.25\linewidth}
         \centering
        \includegraphics[width=\textwidth]{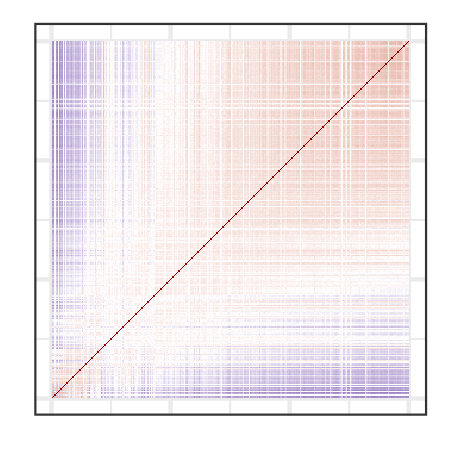}
        \caption{DPM $\sigma = 1$ }
     \end{subfigure}
     \hfill
       \begin{subfigure}[c]{0.13\linewidth}
         \centering
        \includegraphics[width=\textwidth]{figures/sim_gg_heatmap_legend.png}
     \end{subfigure}
     \hfill
     \caption{Ground truth (a-c) and DPM predicted clustering heatmaps (d-f) from the simulation study. Color indicates the probability that two AEs are assigned to the same mixture component (i.e. cluster together). The axes sorted based on the true $\beta$ values in increasing order from left to right.}
    \label{fig:sim_heatmaps}
\end{figure}

We set up simulation studies to compare our proposed DPM model (allowing information sharing between logROR parameters) against a logistic regression with independent normal priors for the logROR parameters (referred to as the IL model), which does not allow information sharing. We investigated two IL models: one has a more informative prior on the logROR parameters (IL-Informative), with precision of 0.1, whereas the other one (IL-Vague) has a smaller precision of 0.01. To mimic the real data and have a distribution of AE counts similar to that from the VAERS data, we took a random sample of 300 intercept terms from the mRNA (target) vs FLU (control) analysis (Section \ref{sec:mrna_flu}) using a logistic regression model. We then simulated the logROR parameters from a mixture of three normal distributions centered at -2.5, 0, and 2.5, with size 50, 150 and 100, respectively. This corresponds to the belief that a majority of the AEs will not be associated with any vaccine, however more AEs will be associated with the target vaccine than with the control vaccine. We also assumed the within-cluster standard deviation to be the same for the three normal distributions, taking  $\sigma = 0.5, 0.8,$ or $1$. As $\sigma$ increases, the clusters become less distinct (as can be seen from the heatmaps in Figures \ref{fig:sig05_gt_htmp}, \ref{fig:sig08_gt_htmp} and \ref{fig:sig1_gt_htmp}) and so we expect the clustering ability of our model to worsen.


We generated 250 datasets for each value of $\sigma$ under the above set-up. All simulated datasets have sparse AE counts; approximately 33\% of the AEs have 0 counts for the control vaccine group. Across the three AE clusters (centered at -2.5, 0, 2.5) between 32-33\%, 11-12\%, and approximately 4.5\% of the AEs have 0 counts for the target vaccine group respectively. 

The precision for $\boldsymbol\alpha_j$ in both the DPM and IL models is set to be $\tau_\alpha = 0.01$. For the DPM model, the remaining hyperparameters are: $\tau_0 = 0.01$, $a_0 = b_0 = 1$, $f_0 = r_\tau = 3$, $r_\lambda = \lambda_0 = 0.03$. 



We compared the models in terms of their DIC (to assess overall model fit), MSE (to assess accuracy in estimating the logROR), as well as the coverage probability of the resulting 95\% posterior credible intervals. We also generated the caterpillar plots to visualize logROR estimates and heatmaps to see how well the clustering structure was recovered in the presence of cluster overlap.

Based on the simulation results in Table \ref{tab:simul_results}, the DPM model has dramatically lower DIC and better MSE compared to the IL-Vague model. The IL-Informative model performs better than the IL-Vague model (the informative prior helps with the sparsity present in the data), but its performance is still worse than the DPM model. All models obtain near nominal coverage for the 95\% posterior credible intervals in the studied cases. We also fit the DPM model using more informative priors and found the model performance did not change, suggesting that the model behavior is robust to different hyperparameter values.

\begin{figure}[ht]
    \centering 
    \includegraphics[width=0.8\textwidth]{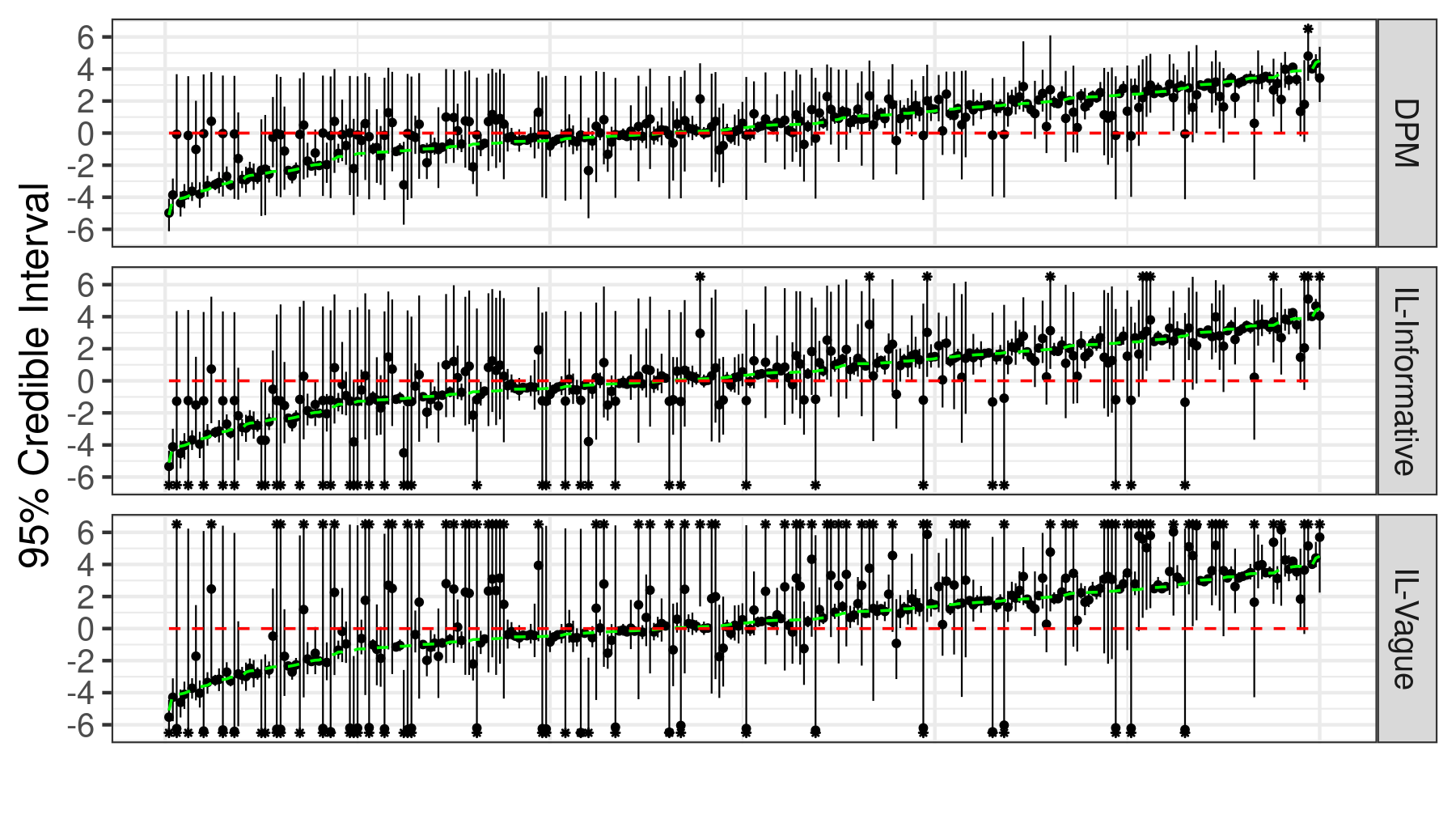}
   \caption{Caterpillar plot from simulation with $\sigma = 0.8$. The dashed green line traces the true $\beta$ values, the dashed red line traces 0, and the 95\% posterior credible intervals are represented by the error bars. Asterisks at the end of the error bars denote credible intervals outside the range of the plot.}
    \label{fig:sim08_caterpillar}
\end{figure}

As shown in Figure \ref{fig:sim08_caterpillar}, the DPM model produces better estimates than the IL models as evidenced by the proximity of the posterior mean estimates to the true values, as well as the narrower confidence intervals while maintaining similar coverage probabilities. For estimates which the DPM model has shrunk towards 0, the uncertainty in the estimates are properly reflected in the larger width of the credible intervals. On the other hand, the IL models do not allow for information sharing between AEs and use the same prior for all logROR estimates. For the IL-Vague model, we observe more extreme estimates and much wider confidence intervals than in the DPM model. For the IL-Informative model, which has a stronger prior to account for the limited information in the sparse data, while the estimates trace the true values better, the resulting confidence intervals are noticeably larger than those from the DPM model. The superior performance in the DPM model is likely due to the information sharing between similar logROR estimates. Using a mixture of normal priors while allowing the number of mixture components to be determined from the data, we are able to induce appropriate shrinkage on the logROR estimates in a data-adaptive way. The caterpillar plots from the $\sigma = 0.5$ and $\sigma = 1$ simulations yielded similar conclusions and can be found in Figures \ref{fig:sim05_cat_plot} and \ref{fig:sim1_cat_plot} in Appendix \ref{appendix:sim_cat_plots}.


The heatmaps in Figure \ref{fig:sim_heatmaps} display the ground truth as well as the learned clustering structure from the DPM model in the simulation cases. For $\sigma = 0.5$, in which there is no overlap between the clusters, the DPM model is able to correctly recover the underlying three-cluster structure. In this case, we attribute the ambiguity in cluster assignment for some of the AEs primarily to the sparsity in the simulated data, especially for the $\beta$ values coming from the extreme clusters. As $\sigma$ increases and the between-cluster heterogeneity decreases, the learned clustering structure becomes less definitive, deviating from the clear block-diagonal structure seen in the $\sigma = 0.5$ simulation. For $\sigma = 0.8$, the DPM model is able to somewhat recover the true cluster structure. We observe more ambiguity in the cluster assignment from the AEs drawn from $\mu_k = 0$, however the extreme clusters are distinct. On the other hand, the DPM model had a much harder time identifying the cluster structure in the $\sigma = 1$ case, in which we observe significantly more overlap between the center cluster and the extreme clusters. However, the overall model performance (based on DIC and MSE) remained better than the IL models, suggesting that the DPM model allowing information borrowing between logROR parameters has advantages over the IL models, even in the absence of a clear underlying cluster structure.


\section{VAERS Data Analysis}

In this paper we investigate safety of the mRNA-based COVID-19 vaccines (Pfizer–BioNTech and Moderna, denoted as ``mRNA" vaccines) and the Johnson \& Johnson–Janssen vaccine (denoted ``JJ" vaccine). We use influenza vaccines (denoted as``FLU") as our reference group. FLU includes inactivated and live attenuated influenza vaccines, labeled as ``FLU3" or ``FLU4," and ``FLUN3" or ``FLUN4" in VAERS. 

We fit our proposed model (denoted ``DPM" model) and compare results to those from IL-Informative (i.e, $\tau = 0.1$, denoted ``IL" model). We did not find much difference in the results when using the IL-Vague model. We are interested in the extent to which the information sharing by way of the DPM prior affects the logROR estimates and model fit. We also applied our proposed negative control and enrichment methods to identify individual AE signals and group-level AE signals in the two studies. As described in the Methods section, an AE is identified as a signal if its NC-adjusted posterior probability is larger than 90\% cutoff and a SOC group $s$ is enriched if the lower bound of the 95\% posterior credible interval of the EOR is greater than 1 and the posterior mean of the EOR  greater than 2.

In all studies, we set a maximum of $K=5$ clusters in the DPM model. We use the same hyperparameter values as in the simulation study for the DPM model. 

We ran three independent MCMC chains for each model. For each chain, after a burn-in of 15,000 iterations we collected one MCMC sample every 5 iterations until we obtained 1,000 MCMC samples. For model convergence, >99\% of the Gelman-Rubin statistics for the $\beta$ coefficients have $R_c$ < 1.2 for the DPM and IL model MCMC chains.

\subsection{Data Preprocessing}
We conducted our analysis using VAERS data from September 2016 to December 2022. All VAERS records for subjects less than 18 years old were omitted from the analysis. All VAERS records for COVID-19 vaccines prior to March 16, 2020 were excluded as these occurred prior to the start of COVID-19 vaccine clinical trials. Finally, all VAERS records indicating administration of both a COVID-19 and non-COVID-19 vaccine were removed as we would not be able to differentiate which vaccine was linked to the AE in question.

 AEs with less than five total cases were excluded from the analysis. In addition, we limit our analysis to AEs for which there were both at least one COVID-19 case and at more than 3 FLU vaccine cases. In the former case, AEs which have not been reported with a COVID-19 vaccine are not relevant as we are interested in AEs associated with the COVID-19 vaccine.

We divided subject age into three groups: [18-30), [30-65), and $\geq 65$. With all categorical variables, we aggregate AE reports by gender, age group, and vaccine group into binomial count data for computational efficiency.

In the enrichment analysis, we used the SOC level in MedDRA to define groups of AEs. For AEs belonging to multiple SOC groups---for example ``Menstruation Delayed" belongs to both the ``Reproductive system and breast" as well as ``Endocrine" disorder groups---when computing the EOR for one of these groups we ignore the AEs' membership in other groups in order to not double-count AEs and bias the EOR estimates.

\subsection{Comparing COVID-19 mRNA and FLU vaccines} \label{sec:mrna_flu}

The final dataset contains 737,301 reports (687,236 mRNA and 50,065 FLU) for 1,089 AEs across 24 SOC groups. Both vaccine groups appear to be predominantly female and have similar sex distributions; the proportion of Unknown, Female, and Male subjects were 1.66\%, 71.9\%, and 26.4\% in the FLU group and 1\%, 69\%, and 30\% in the mRNA group respectively. The FLU group is younger than the mRNA group, having an age distribution of 40.4\%, 49.4\%, and 10.2\% for the three age groups compared to 29.0\%, 59.4\%, and 11.5\% for the mRNA group.

The DPM model fit the data better than IL model, obtaining a DIC of 96,242 compared to 96,271 for IL, indicating that appropriate information sharing in the DPM improves model fit. It uses on average 3.3 mixture components to estimate the logROR coefficients, where we define a component as ``used" if it has at least 5\% of the AEs. 

\begin{figure}[ht]
    \centering
    \includegraphics[width=0.8\textwidth]{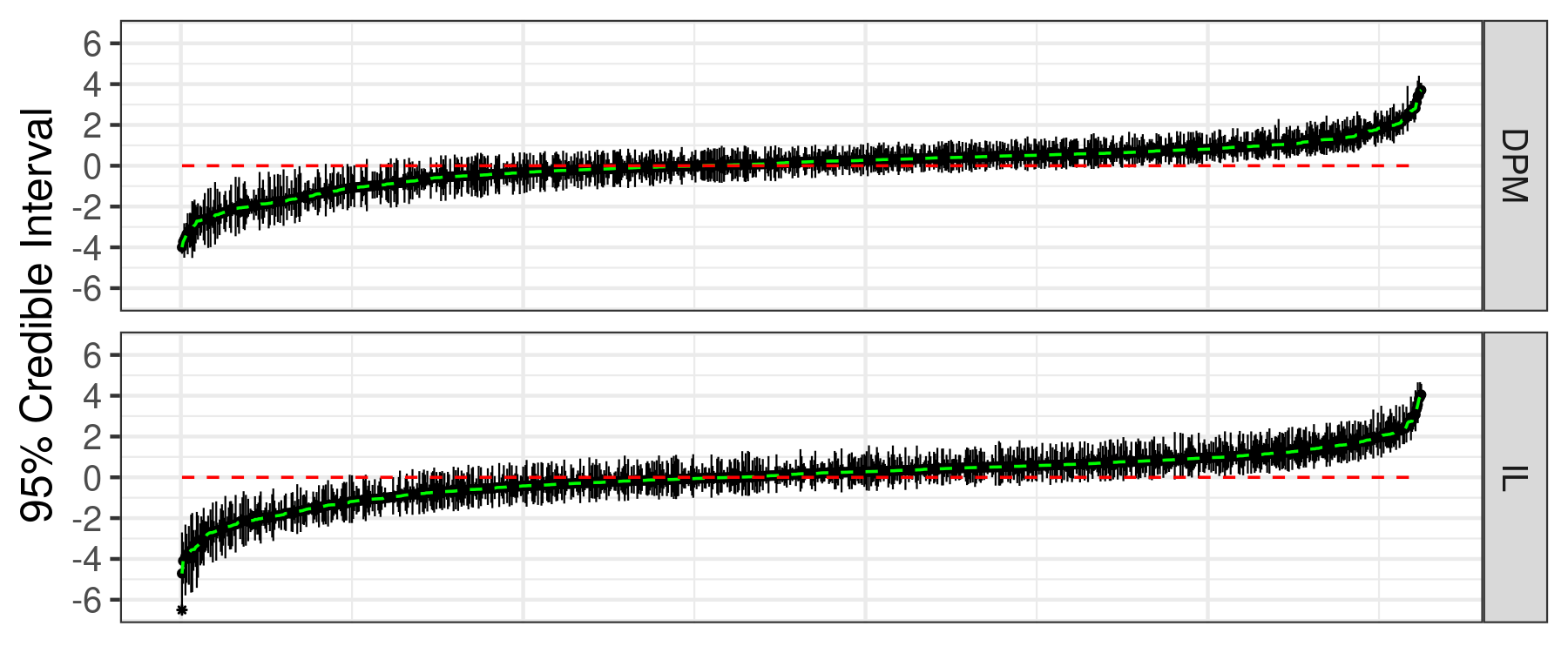}
    \caption{Caterpillar plot in the study for COVID-19 mRNA vaccine, presenting the 95\% posterior credible intervals for the logROR ($\beta$) estimates from both models. The dashed green line shows values of the posterior means and the red dashed line corresponds to $\beta=0$. }
    \label{fig:mrna_caterpillar}
\end{figure}

\begin{figure}[hb]
\centering    
    \label{fig:mrna_soc_enrich_venn}
\hfill
     \begin{subfigure}[c]{0.25\linewidth}
        \includegraphics[width=\textwidth]{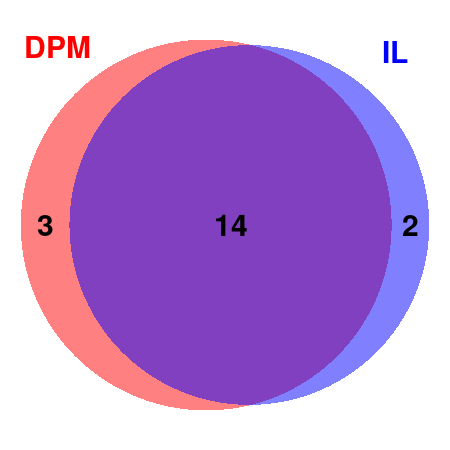}    
        \caption{~}
	\label{fig:mrna_venn}
     \end{subfigure}
    \hfill
     \begin{subfigure}[c]{0.675\textwidth}
         \centering
        \includegraphics[width=\textwidth]{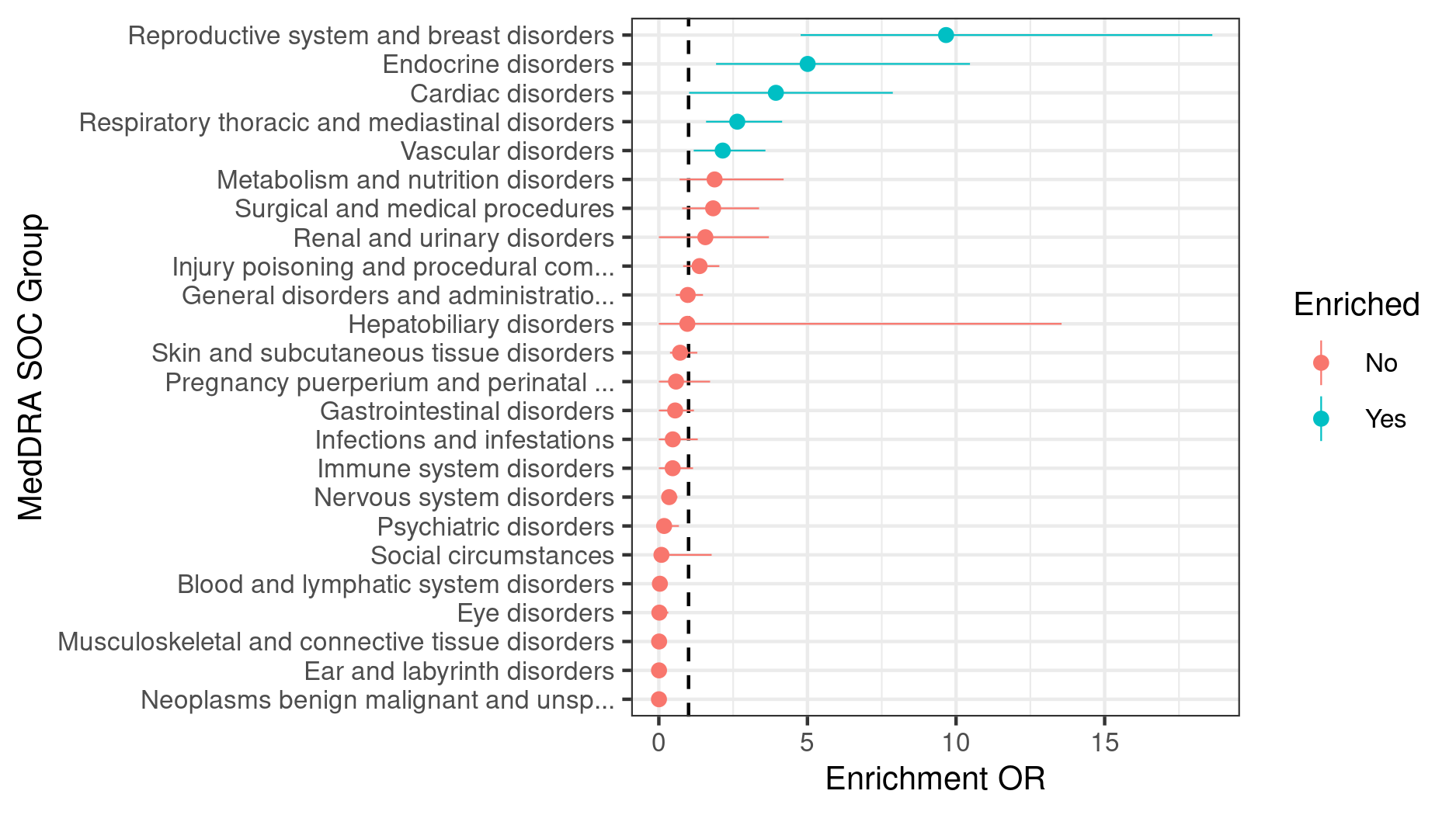}
        \caption{~}
	\label{fig:mrna_soc_enrich}
     \end{subfigure} 
         \caption{(a) Venn diagram comparing the Signal AE overlap between IL (blue) and DPM (red) models and (b) Results from SOC group enrichment analysis for the study of COVID-19 mRNA vaccines.}
\end{figure}

Figure \ref{fig:mrna_caterpillar} shows the credible interval estimates for the DPM and IL models. The IL model credible intervals tend to be wider than those from the DPM model, which is expected from the information sharing in the DPM prior. In addition, the more extreme $\beta$ estimates from the IL model tend to be further from 0 than those from the DPM model, especially for the negative $\beta$ estimates which have a higher incidence of data sparsity and small counts present. 

We identified 17 signal AEs for the mRNA vaccines using the DPM model; see Table \ref{tab:mrna_flu_res} in Appendix \ref{appendix:tables} for logROR estimates and credible intervals as well as the NC-adjusted signal probability. Generally the IL and DPM models identified the same set of signal AEs as shown in the Venn diagram in Figure \ref{fig:mrna_venn}. The differences between the identified signal AEs are that the DPM model identifies ``death," "thrombosis," and ``anticoagulant therapy" while IL identified ``myocarditis" and ``appendicitis."


The identified signal AEs for pulmonary conditions---acute respiratory failure and pulmonary embolism---as well as acute myocardial infarction were shown to be associated with the mRNA COVID-19 vaccines in other studies \cite{miyazato2022multisystem, jabagi2022myocardial}. Similarly the identified reproductive system signal AEs---irregular and delayed menstruation as well as menstrual cycle disorders---and vascular system AEs---deep vein thrombosis and thrombosis---have been linked to the mRNA COVID-19 vaccines \cite{miyazato2022multisystem, edelman2022association, pietri2022p}. Acute kidney injury \cite{d2021minimal}, urticaria \cite{magen2022chronic}, as well as vaccine site pruritus and erythema \cite{doyle2022persistent, cyrenne2021pityriasis, borg2022pfizer} have also been previously reported with the mRNA COVID-19 vaccines.

Other identified AE signals, including anosmia and ageusia (loss of smell and taste respectively), death, anticoagulant therapy, and hypoxia (reduced blood oxygen levels) are associated with COVID-19 infection and/or hospitalization and the higher reporting rates for these AEs might be related to COVID-19 infection which occurred around the time of vaccination. VAERS takes reports for AEs that occur after the administration of a vaccine, regardless whether or not it was caused by the vaccine. Large epidemiological studies are needed to further interrogate these signals and identify a causal mechanism should one be present.

\begin{figure}[ht]
    \centering
    \includegraphics[width=0.8\textwidth]{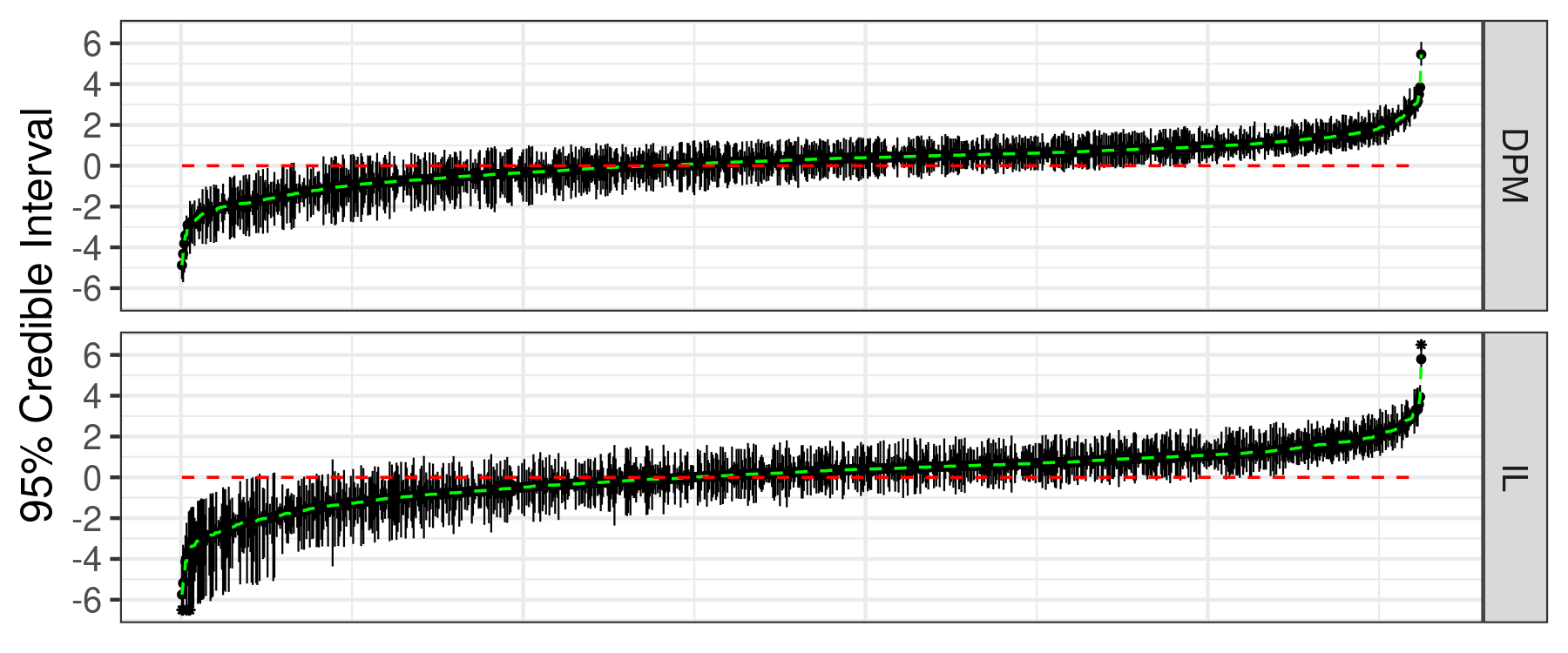}
    \caption{Caterpillar plot in the study for the COVID-19 JJ vaccine, presenting the 95\% posterior credible intervals for the logROR ($\beta$) estimates from both models. The dashed green line shows values of the posterior means and the red dashed line corresponds to $\beta=0$. }
    \label{fig:jj_caterpillar}
\end{figure}

\begin{figure}[hb]
\centering 
\hfill
     \begin{subfigure}[c]{0.25\textwidth}
        \includegraphics[width=\textwidth]{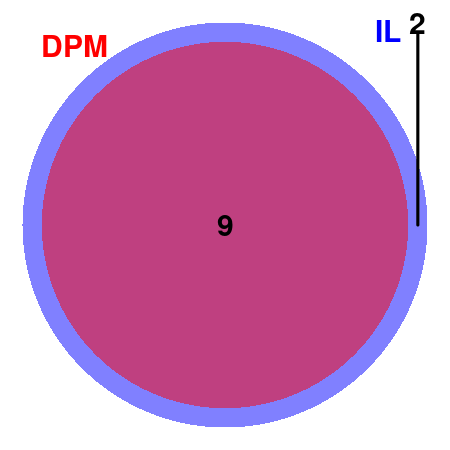}    
        \caption{~}
    \label{fig:jj_venn}
     \end{subfigure}
    \hfill
     \begin{subfigure}[c]{0.675\textwidth}
         \centering
        \includegraphics[width=\textwidth]{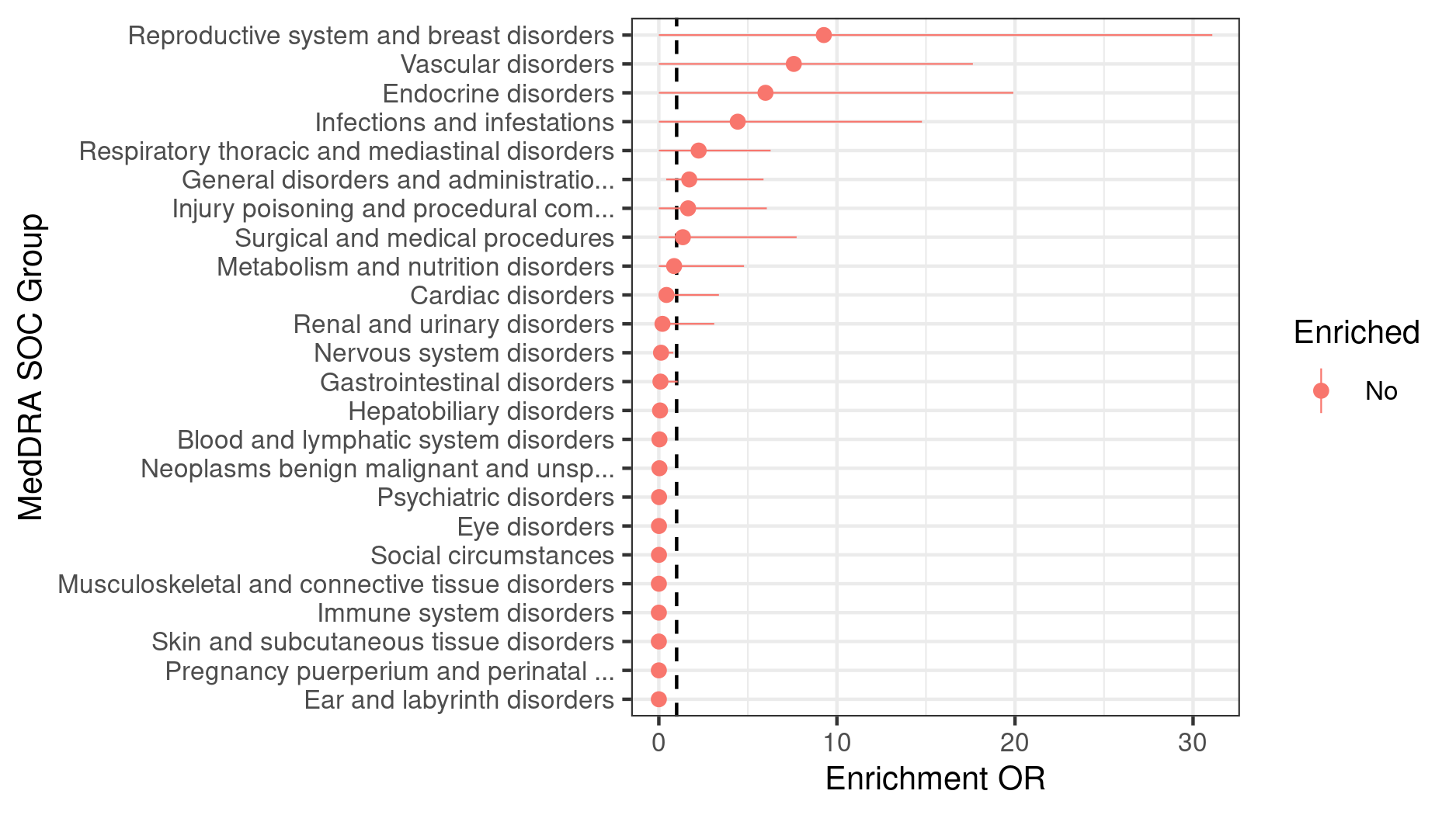}
        \caption{~}
    \label{fig:jj_soc_enrich}
     \end{subfigure}
     \caption{(a) Venn diagram comparing the Signal AE overlap between IL (blue) and DPM (red) models and (b) Results from SOC group enrichment analysis for the study of the COVID-19 JJ vaccine.}
    \label{fig:jj_soc_enrich_venn}
\end{figure}

As shown in Figure \ref{fig:mrna_soc_enrich}, the reproductive system and breast, endocrine, cardiac, respiratory thoracic and mediastinal, as well as vascular disorder groups are significantly enriched for the mRNA COVID-19 vaccines. These results are consistent with the identified signal AEs described above. While individually AEs may not qualify as signals, if enough AEs within a group are close to being signals then this can give an `overall' signal for the group of AEs. The posterior mean EOR values and 95\% credible intervals can be found in Table \ref{tab:mrna_soc_enrich} in Appendix \ref{appendix:soc_tbls}.

\subsection{Comparing COVID-19 Johnson \& Johnson–Janssen and FLU vaccines}

The final dataset contains 106,126 reports (53,901 JJ COVID-19 records and 52,225 FLU) for 1,089 AEs, across 24 SOC groups. Both vaccine groups appear to be predominantly female, however the JJ group has more balanced female/male distribution; the proportion of Unknown, Female, Male subjects were 1.61\%, 71.6\%, and 26.8\% in the FLU group and 0.4\%, 60.4\%, and 39.2\% in the mRNA group respectively. The FLU group is younger than the JJ group, having an age distribution of 40.2\%, 49.6\%, and 10.2\% for the three age groups, respectively, compared to 14.5\%, 67.7\%, and 17.8\% for the JJ group.

The DPM model appears to fit the data better than the IL model, obtaining DIC values of  68,011  and 68,235 respectively.

As in the mRNA vaccine case,  Figure \ref{fig:jj_caterpillar} shows that IL credible intervals are wider than those from the DPM model, with larger differences in width for AEs in the tails of the distribution. The posterior mean logROR estimates appear to be more extreme for the IL model as well.

We identify 9 signal AEs for the JJ vaccine; see Table \ref{tab:jj_flu_res} in Appendix \ref{appendix:tables} for logROR estimates and credible intervals as well as the NC-adjusted signal probability. Generally the IL and DPM models identify the same set of signal AEs, as shown in Figure \ref{fig:jj_venn}, with the differences being the IL model identifying ``thrombectomy" and ``menstrual disorder."

Similar to the mRNA vaccine analysis, a host of vascular AEs---pulmonary embolism, pulmonary thrombosis, deep vein thrombosis, and thrombosis---as well as AEs related to the reproductive system---menstruation irregular, menstrual disorder, and dysmenorrhoea. Hypoxia is again identified as a signal AE which, along with acute respiratory failure, may be related to hospitalization and the patient disease status rather than the disease itself. Further epidemiological studies are needed to understand signal AEs and establish any causal relationships.

We do not observe any significantly enriched SOC groups for the JJ vaccine, which is likely due to the smaller sample size compared to the mRNA analysis. The posterior mean EOR values and 95\% credible intervals can be found in Table \ref{tab:jj_soc_enrich} in Appendix \ref{appendix:soc_tbls}.

\section{Conclusions}

VAERS is an important resource for early identification of vaccine safety issues. It is not however without limitations, namely small counts for rare AEs and reporting bias. We propose a Bayesian model using a Dirichlet Process Mixture (DPM) prior to detect AEs associated with the mRNA-based and Johnson \& Johnson–Janssen COVID-19 vaccines while mitigating the reporting bias. By applying data-adaptive shrinkage for information sharing, we are able to improve the estimates for AEs in the presence of data sparsity. Simulation studies demonstrate the improvements in DIC and MSE from using the proposed DPM model when compared to a baseline model without information sharing. Analysis of VAERS data using our proposed model and negative control procedure identifies previously confirmed AE signals, as well as associated system organ class groups for the AEs. VAERS collects data on any AE following vaccination, whether it is related to the vaccine or not. Therefore, we can not draw any causal conclusion that an AE is caused by a vaccine based on our results. The vaccine safety signals discovered in our real data analyses need to be confirmed by larger studies.

Although this paper focuses on the VAERS database, the proposed methods generally apply to other databases which rely on passive reporting, such as FDA Adverse Events Reporting System (FAERS) and the Adverse Drug Reactions (ADR) database for conducting post-marketing drug safety surveillance.


\begin{center}
\large{\textbf{Supplementary Material}}
\end{center}
\begin{itemize}
    \item \textbf{Supplementary Document A:} JAGS Code for DPM Model
    \item \textbf{Supplementary Document B:} List of VAERS NC AEs
\end{itemize}

\begin{center}
\large{\textbf{Funding}}
\end{center}

Research reported in this publication was supported by the National Institute Of Allergy And Infectious Diseases of the National Institutes of Health under Award Number R01AI158543. The content is solely the responsibility of the authors and does not necessarily represent the official views of the National Institutes of Health.

\bibliographystyle{unsrt}  
\bibliography{DPM_AE}

\newpage
\appendix
\section{Signal AE Tables}\label{appendix:tables}
\begin{table}[ht]
\centering
\begin{tabular}{lrrr}
  \hline
Adverse Event Name & $\beta$ Post. Mean & $\beta$ 95\% Credible Interval & Signal Prob. \\ 
  \hline
  Menstruation irregular & 3.47 & (2.72, 4.41) & 1.00 \\ 
  Menstrual disorder & 3.41 & (2.78, 4.17) & 1.00 \\ 
  Vaccination site pruritus & 3.11 & (2.67, 3.53) & 1.00 \\ 
  Deep vein thrombosis & 2.82 & (2.26, 3.46) & 1.00 \\ 
  Acute myocardial infarction & 2.78 & (2.12, 3.52) & 1.00 \\ 
  Hypoxia & 2.69 & (2.34, 3.06) & 1.00 \\ 
  Acute respiratory failure & 2.61 & (2.25, 3.00) & 1.00 \\ 
  Anosmia & 2.57 & (2.13, 2.96) & 1.00 \\ 
  Vaccination site erythema & 2.31 & (2.07, 2.58) & 1.00 \\ 
  Pulmonary embolism & 2.69 & (2.20, 3.18) & 1.00 \\ 
  Ageusia & 2.33 & (1.96, 2.76) & 0.99 \\ 
  Anticoagulant therapy & 2.11 & (1.79, 2.44) & 0.96 \\ 
  Acute kidney injury & 2.22 & (1.79, 2.64) & 0.96 \\ 
  Mechanical urticaria & 2.43 & (1.69, 3.91) & 0.96 \\ 
  Death & 1.98 & (1.82, 2.14) & 0.94 \\ 
  Menstruation delayed & 2.23 & (1.53, 3.07) & 0.91 \\ 
  Thrombosis & 2.00 & (1.65, 2.42) & 0.91 \\ 
   \hline
\end{tabular}
\caption{Adverse events with at least a 90\% signal probability for the mRNA COVID-19 vaccines when compared to the FLU vaccine. Posterior mean of the logROR and the 95\% credible interval are also presented.}
\label{tab:mrna_flu_res}
\end{table}

\begin{table}[hb]
\centering
\begin{tabular}{lrrr}
  \hline
Adverse Event Name & $\beta$ Post. Mean & $\beta$ 95\% Credible Interval & Signal Prob. \\ 
  \hline
  Vaccine breakthrough infection & 3.84 & (3.41, 4.30) & 1.00 \\ 
  Deep vein thrombosis & 3.49 & (3.08, 3.97) & 1.00 \\ 
  Pulmonary embolism & 3.05 & (2.70, 3.43) & 1.00 \\ 
  Menstruation irregular & 3.16 & (2.64, 3.82) & 1.00 \\ 
  Thrombosis & 2.89 & (2.59, 3.18) & 0.99 \\ 
  Hypoxia & 2.84 & (2.52, 3.18) & 0.99 \\ 
  Acute respiratory failure & 2.79 & (2.47, 3.14) & 0.97 \\ 
  Dysmenorrhoea & 3.03 & (2.34, 3.75) & 0.97 \\ 
  Pulmonary thrombosis & 2.99 & (2.28, 3.87) & 0.94 \\ 
   \hline
\end{tabular}
\caption{Adverse events with at least a 90\% signal probability for the Johnson and Jonson COVID-19 vaccine when compared to the FLU vaccine. Posterior mean of the logROR and the 95\% credible interval are also presented.}
\label{tab:jj_flu_res}
\end{table}

\newpage

\section{SOC Group Enrichment Tables}\label{appendix:soc_tbls}

\begin{table}[ht]
\centering

\begin{tabular}{lrr}
  \hline
Name & EOR & 95\% Credible Interval \\ 
  \hline
Reproductive system and breast disorders & 9.66 & (4.77, 18.62) \\ 
  Endocrine disorders & 5.01 & (1.93, 10.47) \\ 
  Cardiac disorders & 3.94 & (1.02, 7.87) \\ 
  Respiratory thoracic and mediastinal disorders & 2.64 & (1.59, 4.15) \\ 
  Vascular disorders & 2.15 & (1.17, 3.59) \\ 
  Metabolism and nutrition disorders & 1.88 & (0.70, 4.20) \\ 
  Surgical and medical procedures & 1.83 & (0.78, 3.37) \\ 
  Renal and urinary disorders & 1.56 & (0.00, 3.70) \\ 
  Injury poisoning and procedural complications & 1.37 & (0.82, 2.04) \\ 
  General disorders and administration site conditions & 0.97 & (0.57, 1.49) \\ 
  Hepatobiliary disorders & 0.96 & (0.00, 13.55) \\ 
  Skin and subcutaneous tissue disorders & 0.71 & (0.38, 1.30) \\ 
  Pregnancy puerperium and perinatal conditions & 0.58 & (0.00, 1.73) \\ 
  Gastrointestinal disorders & 0.55 & (0.00, 1.18) \\ 
  Infections and infestations & 0.47 & (0.00, 1.31) \\ 
  Immune system disorders & 0.47 & (0.00, 1.15) \\ 
  Nervous system disorders & 0.35 & (0.16, 0.63) \\ 
  Psychiatric disorders & 0.18 & (0.00, 0.68) \\ 
  Social circumstances & 0.09 & (0.00, 1.78) \\ 
  Blood and lymphatic system disorders & 0.04 & (0.00, 0.00) \\ 
  Eye disorders & 0.02 & (0.00, 0.32) \\ 
  Musculoskeletal and connective tissue disorders & 0.01 & (0.00, 0.18) \\ 
  Ear and labyrinth disorders & 0.01 & (0.00, 0.00) \\ 
  Neoplasms benign malignant and unspecified incl cysts and polyps & 0.00 & (0.00, 0.00) \\ 
   \hline
\end{tabular}
\caption{SOC group enrichment OR (EOR) and 95\% credible intervals for COVID-19 mRNA vs FLU vaccine analysis.}
\label{tab:mrna_soc_enrich}

\end{table}

\begin{table}[ht]
\centering
\begin{tabular}{lrr}
  \hline
Name & EOR & 95\% Credible Interval \\ 
  \hline
Reproductive system and breast disorders & 9.27 & (0.00, 31.08) \\ 
  Vascular disorders & 7.58 & (0.00, 17.64) \\ 
  Endocrine disorders & 5.99 & (0.00, 19.91) \\ 
  Infections and infestations & 4.43 & (0.00, 14.77) \\ 
  Respiratory thoracic and mediastinal disorders & 2.24 & (0.00, 6.28) \\ 
  General disorders and administration site conditions & 1.71 & (0.42, 5.88) \\ 
  Injury poisoning and procedural complications & 1.64 & (0.00, 6.06) \\ 
  Surgical and medical procedures & 1.35 & (0.00, 7.74) \\ 
  Metabolism and nutrition disorders & 0.86 & (0.00, 4.78) \\ 
  Cardiac disorders & 0.43 & (0.00, 3.37) \\ 
  Renal and urinary disorders & 0.20 & (0.00, 3.11) \\ 
  Nervous system disorders & 0.12 & (0.00, 0.81) \\ 
  Gastrointestinal disorders & 0.09 & (0.00, 1.06) \\ 
  Hepatobiliary disorders & 0.07 & (0.00, 0.00) \\ 
  Blood and lymphatic system disorders & 0.03 & (0.00, 0.00) \\ 
  Neoplasms benign malignant and unspecified incl cysts and polyps & 0.03 & (0.00, 0.00) \\ 
  Psychiatric disorders & 0.01 & (0.00, 0.00) \\ 
  Eye disorders & 0.01 & (0.00, 0.00) \\ 
  Social circumstances & 0.01 & (0.00, 0.00) \\ 
  Musculoskeletal and connective tissue disorders & 0.00 & (0.00, 0.00) \\ 
  Immune system disorders & 0.00 & (0.00, 0.00) \\ 
  Skin and subcutaneous tissue disorders & 0.00 & (0.00, 0.00) \\ 
  Ear and labyrinth disorders & 0.00 & (0.00, 0.00) \\ 
  Pregnancy puerperium and perinatal conditions & 0.00 & (0.00, 0.00) \\ 
   \hline
\end{tabular}
\caption{SOC group enrichment OR (EOR) and 95\% credible intervals for COVID-19 JJ vs FLU vaccine analysis.}
\label{tab:jj_soc_enrich}
\end{table}

\newpage

\section{Simulation Caterpillar Plots}\label{appendix:sim_cat_plots}

\begin{figure}[ht]
    \centering
    \includegraphics[width=0.8\textwidth]{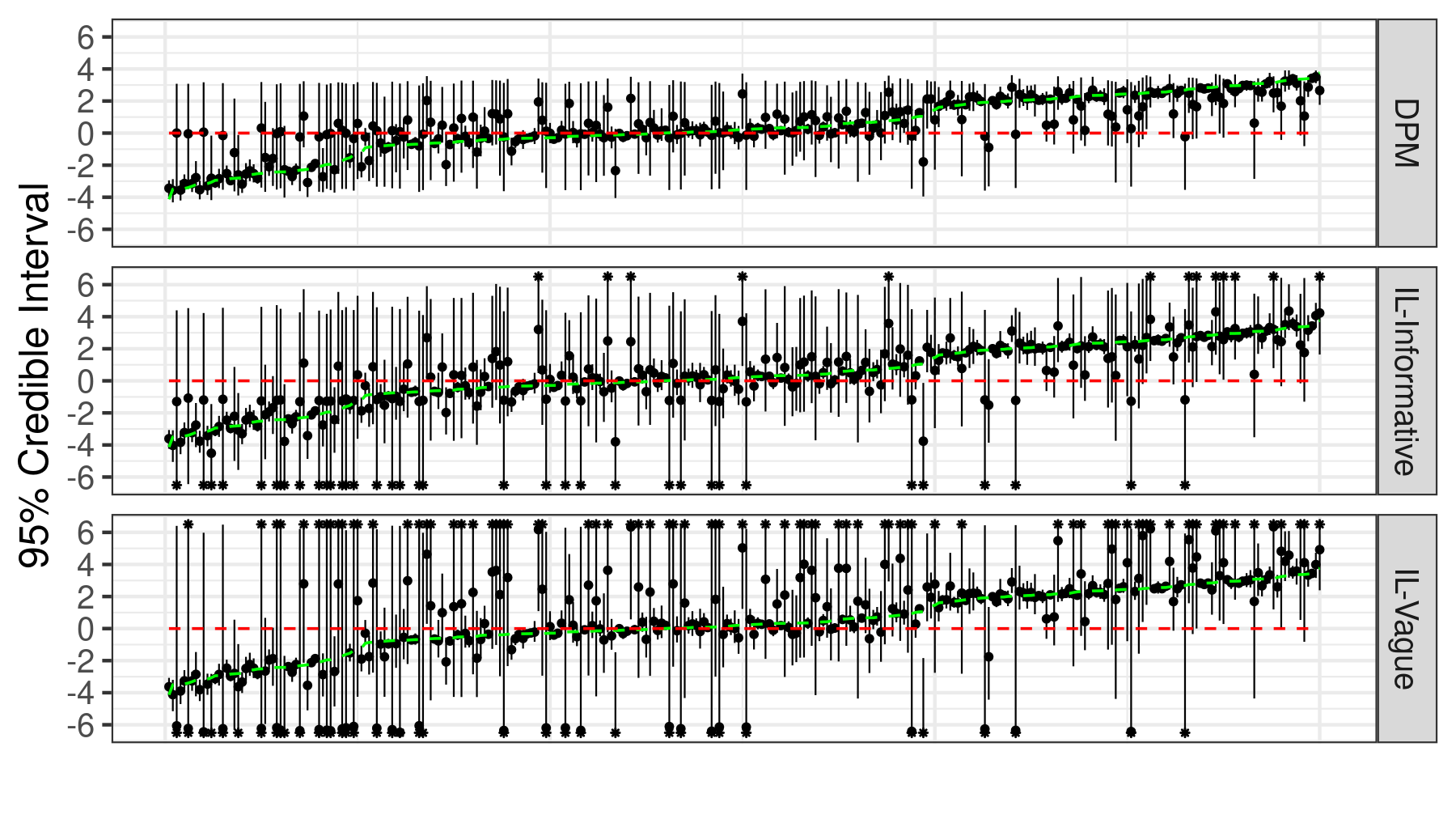}
    \caption{Caterpillar plot from simulation with $\sigma = 0.5$. The dashed green line traces the true $\beta$ values, the dashed red line traces 0, and the 95\% posterior credible intervals are represented by the error bars. Asterisks at the end of the error bars denote credible intervals outside the range of the plot.}
    \label{fig:sim05_cat_plot}
\end{figure}

\begin{figure}[ht]
    \centering
    \includegraphics[width=0.8\textwidth]{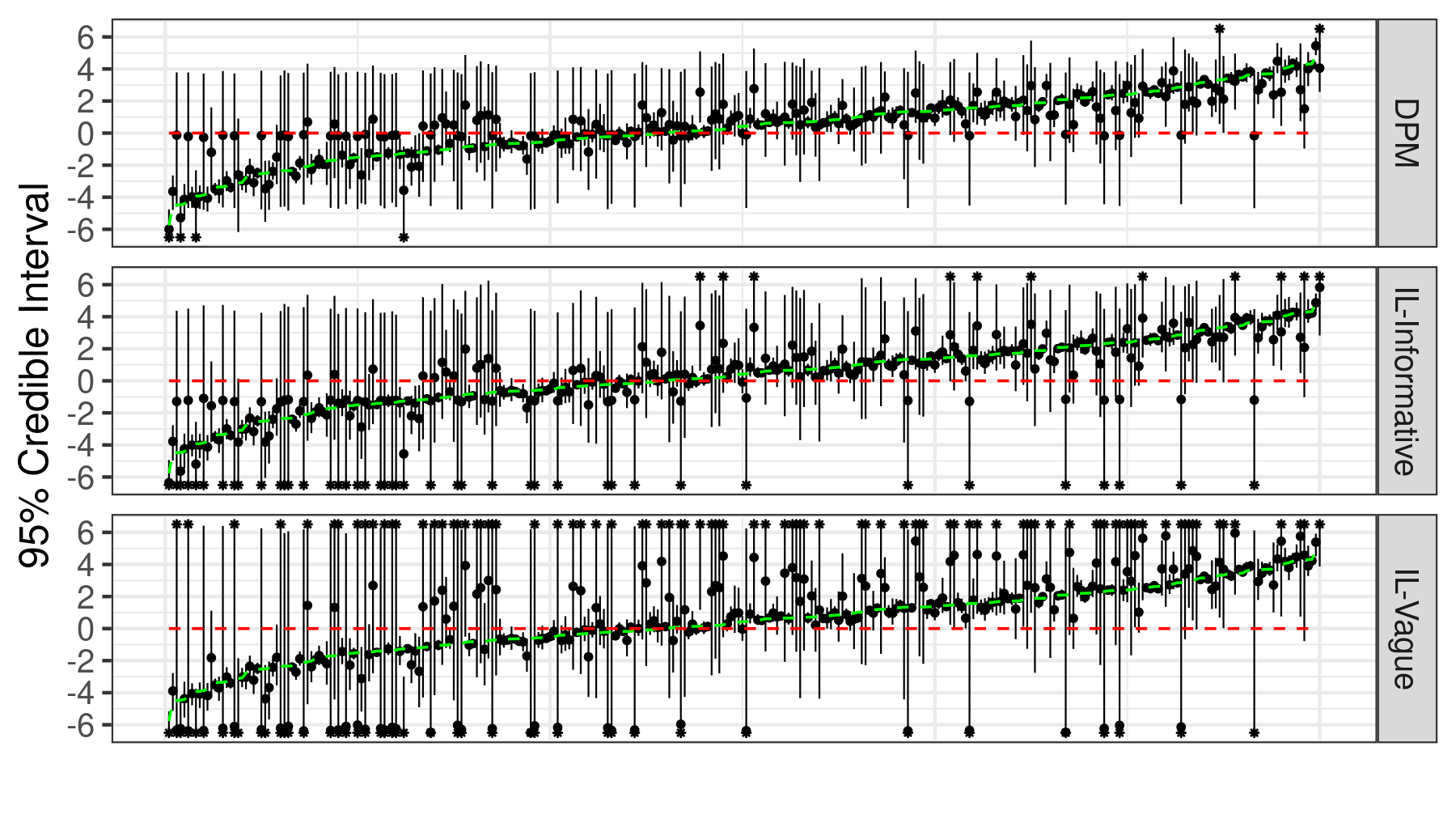}
    \caption{Caterpillar plot from simulation with $\sigma = 1$. The dashed green line traces the true $\beta$ values, the dashed red line traces 0, and the 95\% posterior credible intervals are represented by the error bars. Asterisks at the end of the error bars denote credible intervals outside the range of the plot.}
    \label{fig:sim1_cat_plot}
\end{figure}

\end{document}